\documentclass[]{aa}
\usepackage{graphicx}
\usepackage{txfonts}
\begin{document}
%
%
\newcommand{\tuc}{\rm $J$=1--0}            
\newcommand{\tdu}{\rm $J$=2--1}         
\newcommand{\ttd}{\rm $J$=3--2}         
\newcommand{\doce}{\rm $^{12}$CO}       
\newcommand{\trece}{\rm $^{13}$CO}      
\newcommand{\gsim}{\raisebox{-.4ex}{$\stackrel{>}{\scriptstyle \sim}$}}
\newcommand{\lsim}{\raisebox{-.4ex}{$\stackrel{<}{\scriptstyle \sim}$}}
\newcommand{\psim}{\raisebox{-.4ex}{$\stackrel{\propto}{\scriptstyle \sim}$}}
\newcommand{\kms}{\mbox{km~s$^{-1}$}}
\newcommand{\jyb}{\mbox{Jy~beam$^{-1}$}}
\newcommand{\s}{\mbox{$''$}}
\newcommand{\mloss}{\mbox{$\dot{M}$}}
\newcommand{\my}{\mbox{$M_{\odot}$~yr$^{-1}$}}
\newcommand{\ls}{\mbox{$L_{\odot}$}}
\newcommand{\ms}{\mbox{$M_{\odot}$}}
\newcommand{\mm}{\mbox{$\mu$m}}
\def\arcdeg{\hbox{$^\circ$}}
\newcommand{\secp}{\mbox{\rlap{.}$''$}}
\newcommand{\secs}{\mbox{\rlap{.}$^{\rm s}$}}
\newcommand{\um}{\mbox{$\mu$m}}
\newcommand{\h}{$^{\rm h}$}
\newcommand{\m}{$^{\rm m}$} 
\newcommand{\irc}{IRC\,+10420}         
\newcommand{\afg}{AFGL\,2343}         
\newcommand{\nueve}{$^{29}$SiO}
\title{ The structure and chemistry of the massive shell around AFGL\,2343: \nueve\ and HCN as tracers of high-excitation regions \thanks{ Based on observations carried out
with the IRAM Plateau de Bure interferometer. IRAM is supported by INSU/CNRS (France), MPG (Germany)
and IGN (Spain).}  }
\titlerunning{Structure and chemistry in the CSE around AFGL\,2343}
%
%
   \author{
          G. Quintana-Lacaci \inst{1}
          \and
          V. Bujarrabal \inst{1}
          \and
          A. Castro-Carrizo \inst{2}
          }
%
   \offprints{g.quintana@oan.es}
   \institute{
     Observatorio Astron\'omico Nacional (IGN), Apdo. 112, 
                                   E-28803 Alcal\'a de Henares, Spain \\
              \email{(g.quintana,v.bujarrabal)@oan.es}
         \and
     Institut de RadioAstronomie Millim\'etrique, 300 rue de la Piscine,  
                                   38406 Saint Martin d'H\`eres, France \\
              \email{ccarrizo@iram.fr}
     }

\date{Received 2008 / Accepted }

 
  \abstract 
 {The yellow hypergiant stars (YHGs) are very massive objects that are expected to pass through periods of intense mass loss during their evolution. Despite of this, massive circumstellar envelopes have been found only in two of them, \irc\ and \afg.}
  {The envelopes around these objects and the processes that form them are poorly known. We aim to study the structure, dynamics and chemistry of the envelope around \afg.}
  {We have obtained interferometric maps of the rotational lines \nueve\ $J$= 2--1, HCN $J$= 1--0 and SO $J_K$= 2$_2$--1$_1$ towards \afg.  We have used an LVG excitation model to analyze the new observations and some previously published line profiles of \afg.}
  {The analysis of the observational data and the fitting results show the presence of a thin, hot and dense component within the previously identified CO shell. This component can be associated          with recently shocked gas, { but it could also be due to a phase of extremely copious mass loss}. We suggest that this shell is the responsible for the whole \nueve\ emission and significantly contributes to the HCN emission. The presence of such a dense shell rich in SiO can be related with that previously found for \irc, which was also suggested to result from a shock. This may be a common feature in the evolution of these stars, as a consequence of the episodic mass loss periods that they pass during their evolution. We present new results for the mass loss pattern, the total mass of the circumstellar envelope and the molecular abundances of some species in \afg. }
  {}

   \keywords{(Stars:) circumstellar matter -- (Stars:) supergiants -- Stars: AGB and post-AGB
     -- Radio lines: stars -- Stars: individual:
    AFGL\,2343}

   \maketitle

\section{Introduction}

The yellow hypergiants stars (YHGs) are amongst the most luminous (5.3
$\leq$ log$L$[\ls] $\leq$ 5.9) and massive ($M_{\rm init} \sim$ 20 \ms) stars in the sky 
(see, as general references, de Jager 1998, Jones et al$.$ 1993, Humphreys 1991).
These stars are thought to be post-red supergiant objects evolving bluewards in the 
HR diagram, but the details of such an evolution are still poorly known. In at least a few of them, the stellar temperature is rapidly increasing. For example, the spectral 
type of the YHG IRC\,+10420 has changed from F8Ia to A5Ia in just 20 yr 
(Oudmaijer et al. 1996, Oudmaijer 1998, Klochkova et al.\ 1997). Humphreys et al.
(2002) showed, however, that the wind in this source is optically thick, suggesting 
that the apparent spectral type changes can be due to variations in the wind rather 
than to interior evolution. Smith et al. (2004) also presented numerical simulations of such an effect. 

Although it is thought that, during the red and yellow phases, 
these heavy stars eject up to one half of their initial mass 
(e.g.\ Maeder \& Meynet 1988, de Jager 1998), only two YHGs, IRC\,+10420 (= IRAS\,19244+1115) and AFGL\,2343 (= IRAS\,19114+0002 = HD179821), show 
very heavy circumstellar envelopes (CSEs). Those CSEs are detected in molecular line emission, dust-scattered light and IR emission (see Hawkins et al$.$ 1995, Meixner et al$.$ 1999, Humphreys et al$.$ 1997, Bujarrabal et al$.$ 2001, Castro-Carrizo et al.\ 2001, Quintana-Lacaci et al. 2007). 
The Hipparcos distance for AFGL\,2343 is 5.6 kpc, { but the uncertainty of this measurement is high, comparable to the parallax value itself. 
The poor} accuracy of this measurement led Josselin \& L\`ebre (2001) to propose it to be a young planetary nebula at a distance of about 1\,kpc. On the contrary, Hawkins et al. (1995) and Jura et al. (2001) argue that AFGL\,2343 is a massive star at a distance of 6\,kpc, in 
agreement with the kinematic distance estimate. Recent results based on the similarities of its CSE with
that of IRC\,+10420 confirm that AFGL\,2343 is a real yellow hypergiant star (Castro-Carrizo et al. 2007, Quintana-Lacaci et al. 2007). Consistently, here we adopt a distance of { 6\,kpc.}

Castro-Carrizo et al. (2007) (hereafter CC07) showed that the heavy envelope surrounding \afg\ was formed during different periods of mass loss, which at present is however very low. Those periods lasted typically a thousand years, the total time for the formation of the envelope being $\sim$4500 years. 
The properties of the CSEs around AFGL\,2343 and IRC\,+10420 are compatible with mass loss driven by radiation pressure.

Quintana-Lacaci et al. (2007) (hereafter QL07) studied the chemistry and abundances of many molecular species in \afg. The chemistry was found to be similar to that of O-rich AGB stars, except for a general underabundance of molecules other than CO. Note, however, that the lack of information on the spatial distribution of the line-emitting regions severely hampered the abundance estimates.


\section{Observations}

We have obtained interferometric maps with the Plateau de Bure interferometer (PdBI, France) of the molecular transitions HCN $J$=1--0, $^{29}$SiO $J$=2--1 and SO $J_K$=2$_2$--1$_1$ in the yellow hypergiant AFGL\,2343.
The interferometer consists of 6 antennas of 15\,m  in diameter with dual-polarization single-band SIS heterodyne receivers. Observations were performed in configuration 6Bq (in March, 2007) and 6Cq (in April, 2007). The projected baselines ranged from 16\,m to 452\,m. AFGL\,2343 was observed at coordinates 19\h26\m48$\secs$10 +11\degr21\arcmin17$\secp$0. MWC\,349 and 3C\,273 were observed to calibrate absolute flux. 1923+210, 1749+096, and J2025-075 were used
to calibrate the evolution of visibility amplitudes and phases along time. The accuracy of the flux calibration is within 10\%
at 3\,mm. The calibration and data analysis were performed in the standard way using the GILDAS\footnote{See {\tt http://www.iram.fr/IRAMFR/GILDAS} for more information about the GILDAS software.} software package. Imaging and cleaning was performed by using natural weights and verifying that the flux contained in the
found CLEAN components corresponds to that seen in the uv-tables.
Neither halos nor elongations were seen in the continuum maps of the phase
calibrators. It is therefore not expected to have any spurious contribution from the calibration above the dynamic range.

Final maps are shown in Figs.\,1--\,3. By comparison with single-dish data (QL07)
we find that the amount of flux filtered out by the interferometer in our observations of \nueve\ and HCN is small, $\lsim$ 10\%.

   \begin{figure*}
   \centering
   \includegraphics[width=6cm,angle=-90]{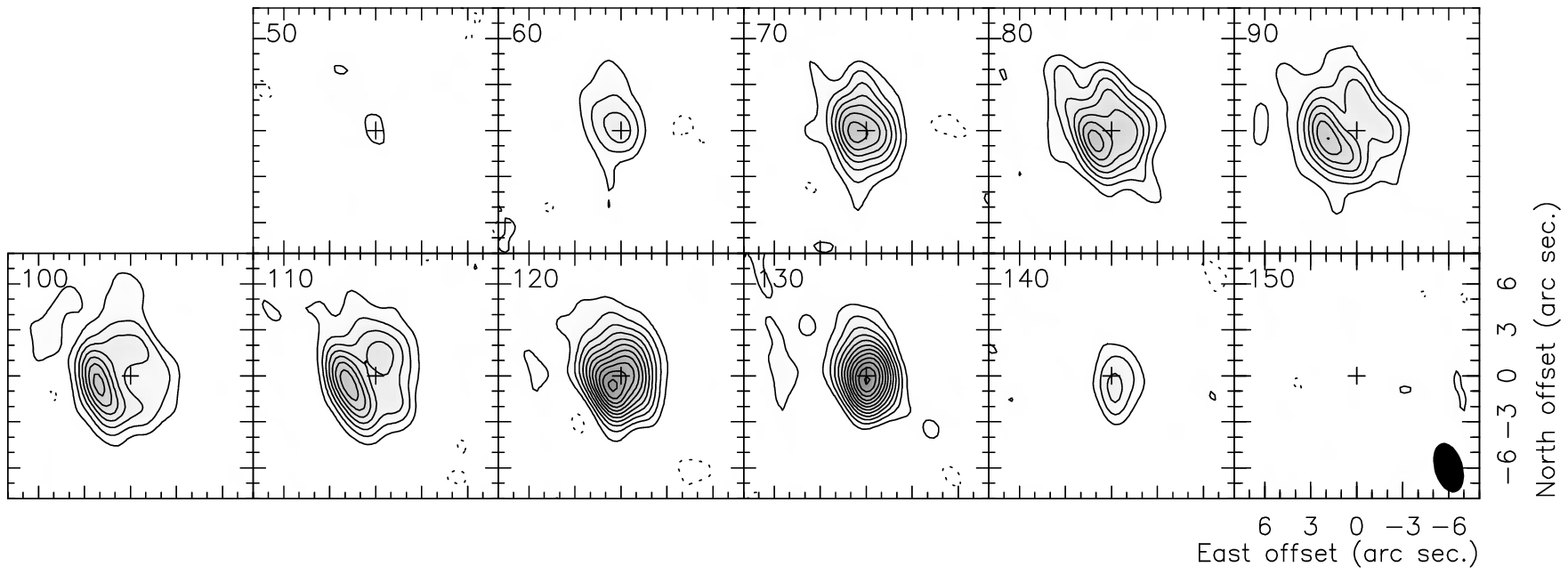}
      \caption{
PdBI maps of the \nueve\ $J$ = 2--1 emission in AFGL\,2343. LSR velocities (in \kms) are indicated in the upper left corner of each box. The first contour and the level step are at 1.8 $\times$  $\sigma$ = 2 mJy/beam. 
The CLEAN beam (at half-power level), of size 3\secp27 $\times$ 1\secp84 ($FWHM$) and PA 15\degr, is drawn in the bottom right corner of the last panel.
             }
         \label{map1}
   \end{figure*}

   \begin{figure*}
   \centering
   \includegraphics[width=8cm,angle=-90]{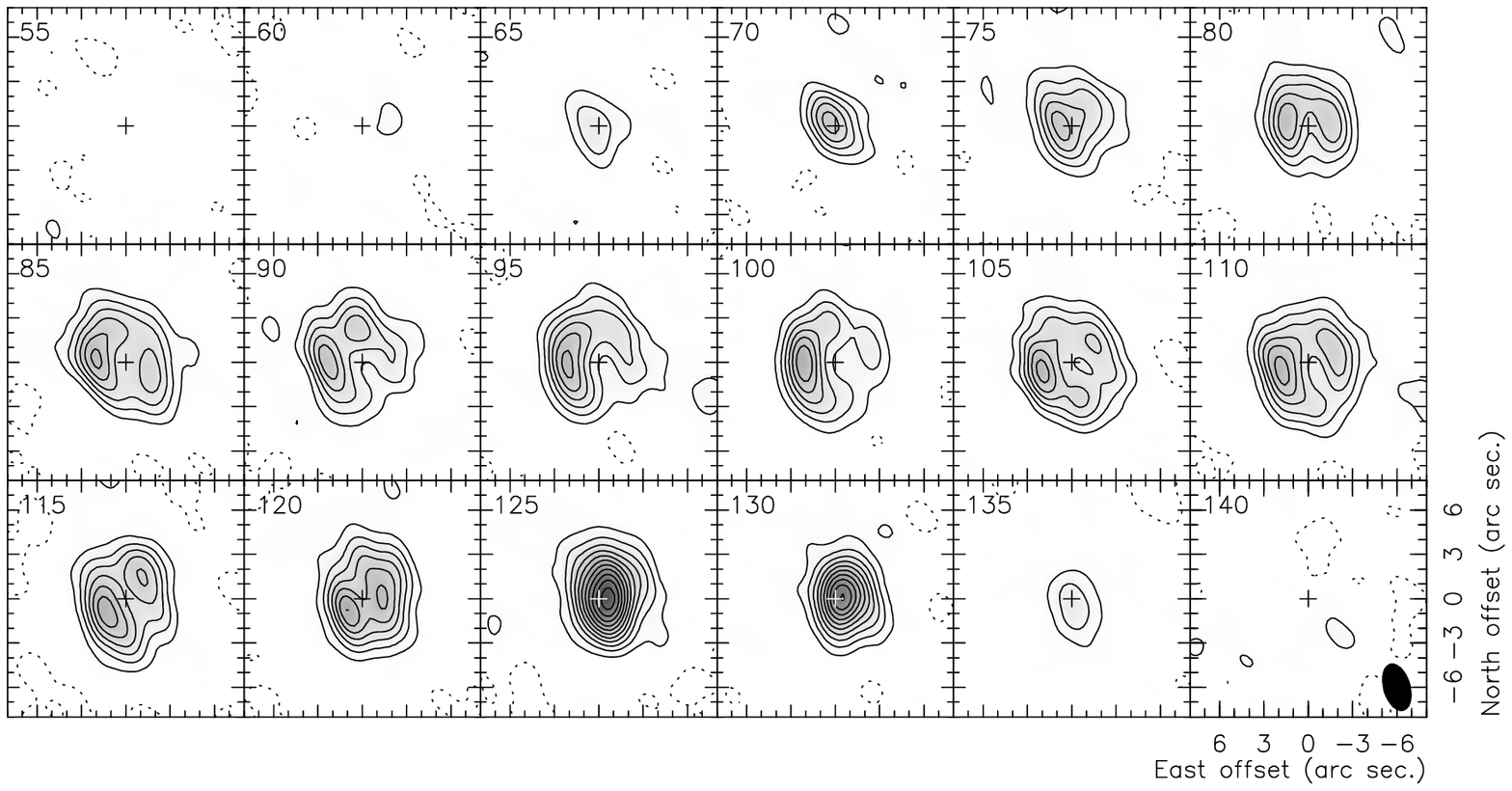}
      \caption{
PdBI maps of the HCN $J$ = 1--0 emission in AFGL\,2343. LSR velocities (in \kms) are indicated in the upper left corner of each box. The first contour and the level step are at 3 $\times$ $\sigma$ = 5 mJy/beam. 
The CLEAN beam (at half-power level), of size 3\secp31 $\times$ 1\secp85 ($FWHM$) and PA -164\degr, is drawn in the bottom right corner of the last panel.
              }
         \label{map2}
   \end{figure*}

   \begin{figure*}
   \centering
   \includegraphics[width=6cm,angle=-90]{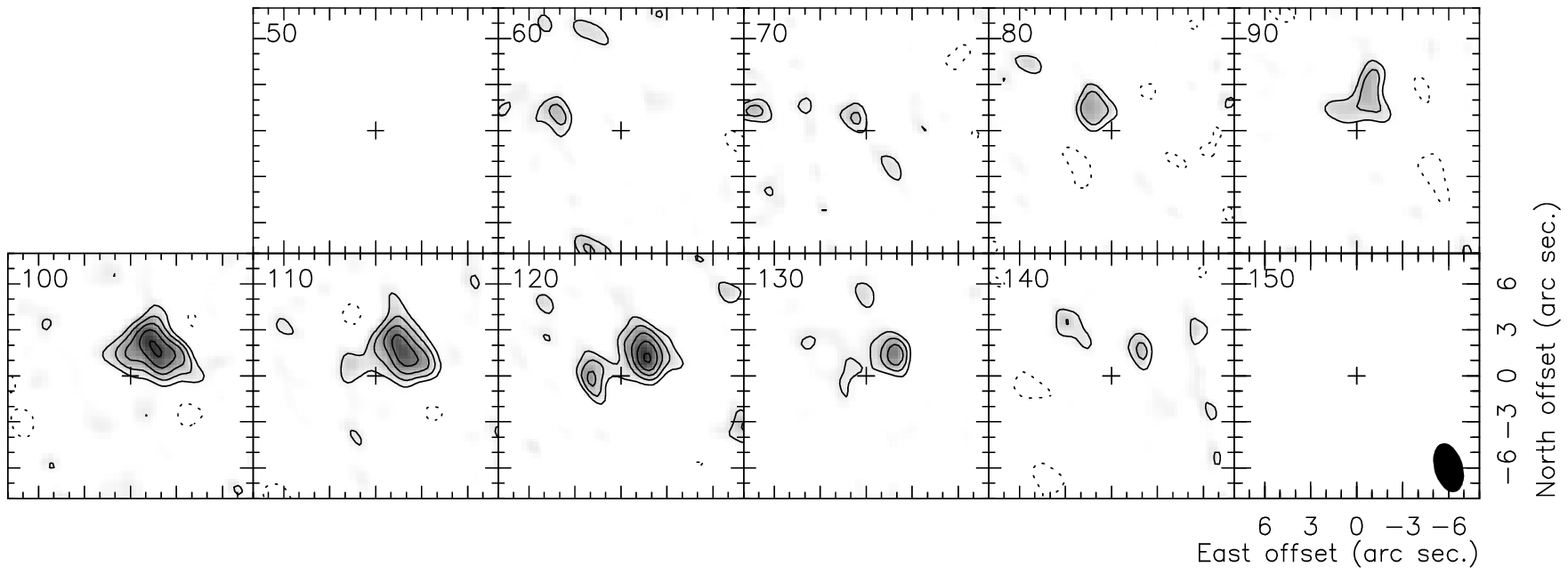}
      \caption{
PdBI maps of the SO $J_K$ = 2$_2$--1$_1$ emission in AFGL\,2343. LSR velocities (in \kms) are indicated in the upper left corner of each box. The first contour and the level step are at 0.9 $\times$ $\sigma$ = 1 mJy/beam. 
The CLEAN beam (at half-power level), of size 3\secp26 $\times$ 1\secp84 ($FWHM$) and PA 16\degr, is drawn in the bottom right corner of the last panel.
              }
         \label{map3}
   \end{figure*}


\section{Analysis of the interferometric maps}
\subsection{Description of the molecular emission}

The maps presented in the previous section (Figs.\,1--3) show that the emission of HCN and \nueve\ is mainly spherical, presenting a central hole similar to that of the CO maps (CC07).

Our maps present however some departures from circular symmetry. The central velocity channels (from 70\,\kms\ to 110\,\kms) of the $^{29}$SiO map show that the emission peaks mainly at the east of the shell. In the HCN maps this is also seen, but is less evident. 
We also find a maxima for both lines at positive velocity (at 125\,\kms\ for HCN and at 130\,\kms\ for \nueve), slightly shifted towards the west. { Similar deviations from spherical symmetry were also observed in OH maser emission
(Gledhill et al. 2001).}

The asymmetries found for HCN $J$=1--0 and $^{29}$SiO $J$=2--1 are not dominant in the overall emission maps, as it can be seen in the integrated emission map of HCN (Fig.\,4). 
The intense peak observed towards the east at the central velocities is compensated by a very intense peak at 125\,\kms\ placed towards the west.
When comparing the profiles of Fig.\,4, corresponding to the easternmost and westernmost emission, we see an excess at central velocities for the former, and an excess for the latter at positive velocity. { The value of the integrated emission is similar for both sides of the integrated map. The lack of emission at central velocities for the westernmost region is compensated by the excess at high velocity. Indeed,} the integrated profile areas of the two excesses are equal ($\sim$ 0.4\,K\,$\kms$).
{ This can be interpreted showing that the spherical symmetry and isotropy of the whole shell is altered by the presence of some regions within it, which strongly emit in these lines, one of them placed towards the east and the other showing a particularly high LSR velocity.}


   \begin{figure*}
   \centering
   \includegraphics[width=4cm,angle=-90]{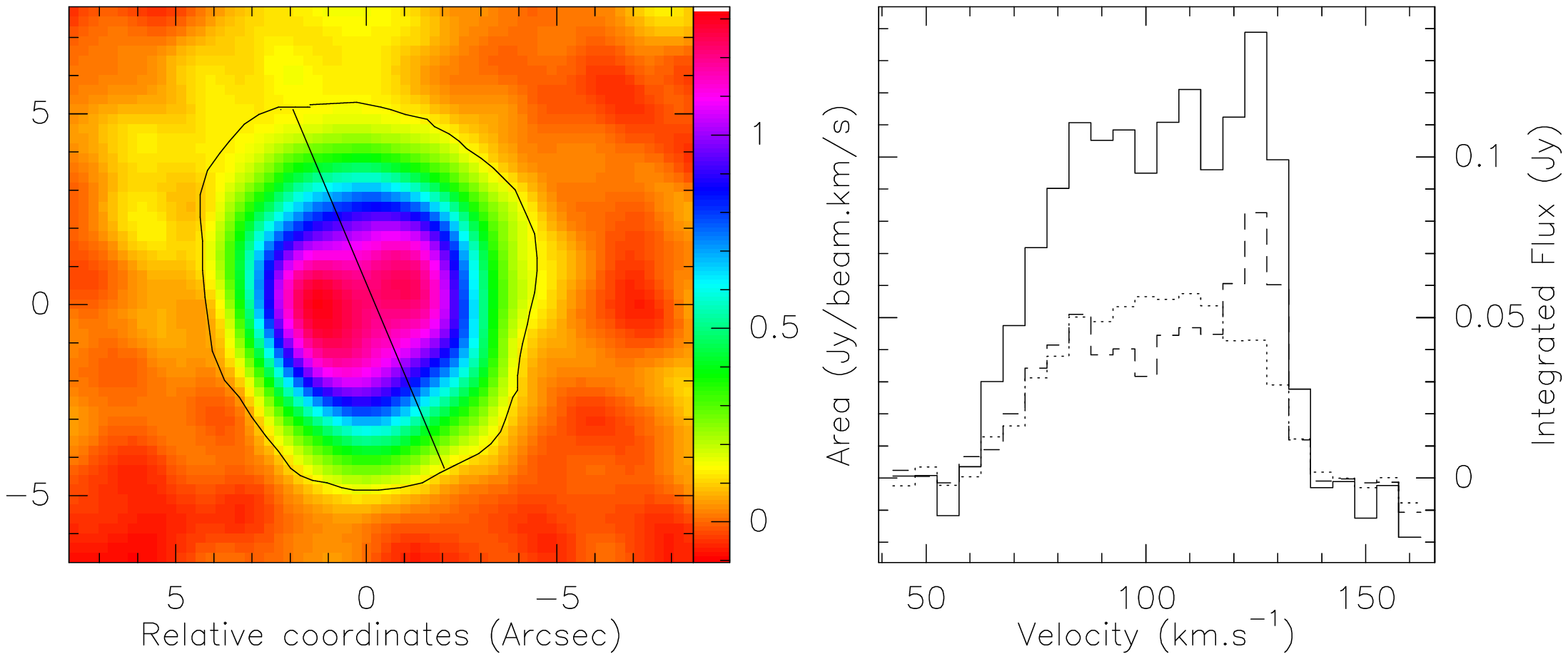}
      \caption{$\it Left$: Integrated HCN $J$=1--0 map. This map is divided in two halves, corresponding to the emission from the east (left) and from the west (right).
$Right$: Spectra comparison of the total flux (solid line) with the emission from the west (dashed line) and from the east (pointed line). 
              }
         \label{compo}
   \end{figure*}

The molecular profiles obtained by QL07 for \nueve\ and HCN also show this narrow emission peak at positive velocity (at 125\,\kms). Note that this emission peak in the profiles of \nueve\ and HCN is relatively more intense in the high-J transitions.{  The ratio between the emission peak at high velocity and the intensity at the center of the line is $\sim$3.5 for \nueve\ $J$=5--4 and $\sim$3.3 for HCN $J$=3--2, while for \nueve\ $J$=2--1 and HCN $J$=1--0 this peak is hardly noticeable. We also find that, for $^{28}$SiO, this ratio is $\sim$1.4, $\sim$1.6 and $\sim$2.6 for transitions $J$=2--1, $J$=3--2 and $J$=5--4, respectively. The relatively high intensity of the positive-velocity peak for high J values indicates that this emission comes from a relatively hot region.}

{ Note also that, in shells around evolved stars, SiO is only present in the hot central regions,
in which the grains are still not completely formed, or in recently shocked regions (e.g. S\'anchez-Contreras et al. 1997). }

{ In view of the anomalous velocities and high excitation characteristics of the above discussed emissions,} we suggest that both the maxima, that detected in the east part of the envelope and that detected at positive velocity, come from a high-excitation region, probably not a complete shell, moving outward faster than the rest of the shell. Because of its relatively high excitation, velocity, and SiO abundance, this shell could result from the passage of a shock front. { Indeed, this is compatible with the suggestion made by Gledhill et al. (2001) that shocks produced by a collimated outflow are the responsible for the asymmetries found in OH emission. However, it is also possible that this shell is due to an extreme mass-loss period, like that observed recently for the YHG $\rho$ Cas (Lobel et al. 2003).}
Therefore, a new component, a high-excitation shell (hereafter HE shell) needs to be added to those obtained by our CO analysis (CC07) to reproduce this new feature; see Sect.\,4.2 for further discussion.


In the case of SO $J_K=2_2-1_1$ the signal-to-noise ratio is low, leading to maps where only the dominant emission peaks are detected (see Fig.\,3). 
\subsection{Main molecular reservoir extension}

   \begin{figure*}
   \centering
   \includegraphics[width=4cm,angle=-90]{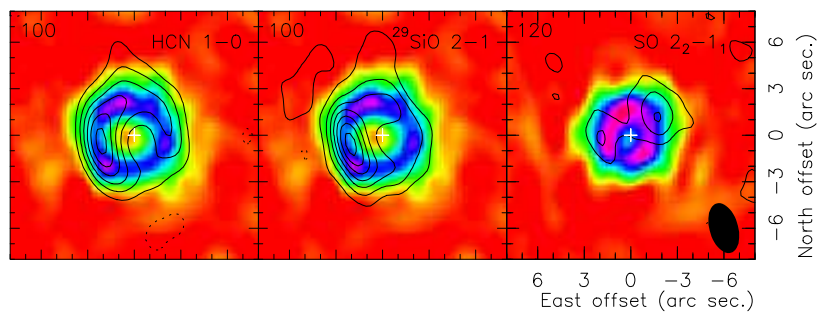}
      \caption{HCN $J$=1--0, $^{29}$SiO $J$=2--1 and SO $J_K$=2$_2$--1$_1$ contour maps 
superposed to $^{12}$CO $J$=2--1 emission (color scale) for the central velocity channel of AFGL\,2343. { In the third panel we show the channel corresponding with 120\,\kms for SO and CO, since at this velocity the extent of the former is better traced.}
              }
         \label{comp}
   \end{figure*}



In Fig.\,5 we present the contour map of $^{29}$SiO, HCN and SO at the central velocity superposed to the CO distribution { at the same velocity} by CC07 (in the case of SO we used the channel at $V_{\rm LSR}=\,120\kms$, which shows better the extent of the emission of this molecule).
This figure shows that the emissions of the different molecules come from similar regions. 
In the case of $^{29}$SiO, the peaks seem closer to the center than the CO ring-like structure, and therefore the \nueve\ emission probably comes from relatively inner 
layers of the CO dense shell found by CC07. 
Note that QL07 assumed that the densest shell of the CO envelope was the responsible for the emission of other molecules than CO. From the results we show here, this assumption was rather correct for most molecules (Sect. 5) except for the presence of the HE shell. The HE shell, that must be included in the modeling to account for the above mentioned peculiarities (see Sect.\,3.1), is therefore located within the dense shell found in CO.

The presence of a new component 
implies changes in the values of the molecular abundances deduced by QL07. We estimate new molecular abundances taking into account the new results derived by including the HE shell (Sect.\,5).


\section{Molecular emission model}
\subsection{Description of the excitation code}

We have modeled the $^{29}$SiO $J$ =2--1 
and HCN $J$ = 1--0 
emission of AFGL\,2343 using a method similar to that described by CC07 and Teyssier et al. (2006). We have used a standard { Large Velocity Gradient (LVG)} code to  
obtain the level population and excitation temperatures. This model takes into account collisional excitation as well as radiative excitation. We have included the collision coefficients from Turner et al. (1992) for $^{29}$SiO, and Green \& Thaddeus (1974) for HCN, and its extrapolation to higher temperatures taken from the {\it Leiden Atomic and Molecular Database}\footnote{\tt http://www.strw.leidenuniv.nl/$\sim$moldata}(LAMDA; see Sch\"oier et al. 2005). This extrapolation has little effect on our calculations, since the temperatures found in AFGL\,2343 are moderate. Only the rotational transitions of the fundamental vibrational level are taken into account in most of our simulations, since 
we have checked that including higher vibrational levels does not significantly change the results.
The radiative excitation does not play an important role in our case: we have also checked that the IR and the mm-wave millimeter emission do not significantly affect the populations of the levels.

We assume spherical symmetry in our calculations, since the circumstellar envelope around AFGL\,2343 is mainly spherical. The density, $n$, of the circumstellar envelope at a given radius, $r$, is determined from the mass-loss rate, $\mloss$, and the expansion velocity, $V_{\rm exp}$, assuming $n=\mloss/(4\pi r^2V_{\rm exp})$. The temperature is defined by adopting a { power law}, as usually assumed for AGB circumstellar envelopes and found to be adequate for YHGs by CC07: $T_{\rm k}(r)=T(r_o)\times(r/r_o)^{-\alpha_t}+ T_{\rm min}$. The local velocity dispersion is described by using a Gaussian function for a standard deviation $\sigma_{\rm tur}$, due to turbulent movements.

The level excitation equations are solved in a finite number of radii, with an increasing separation between the points.
However, to solve the radiative transfer equations in a thin region within a wide envelope the distribution of points of the model cited above is not suitable. To model such a thin region we introduce a tighter distribution of points
in the region of interest, to enhance the number of calculations made in this thin shell and to obtain a realistic description of the molecular excitation in it. These results are interpolated to estimate the conditions at any distance.

From the excitation results of the LVG calculations, we obtain the brightness distribution in the plane of the sky for each velocity channel. This brightness distribution is calculated by solving the radiative transfer equations along the line of sight for different impact parameters. Note that, as we have spherical symmetry, the brightness distribution will show circular symmetry. The brightness distribution is convolved with the Gaussian synthetic beam obtained for each interferometric map, yielding images of main-beam Rayleigh-Jeans-equivalent temperatures. The beam shape is elliptical in our maps, so this convolution will break the circular symmetry of the brightness distribution. 
Since the circumstellar shell in AFGL\,2343 is relatively thin, the convolution with the beam results in two opposite maxima, both elongated in the direction of the beam major axis

We have also used this code to reproduce the line profiles of $^{29}$SiO, $J$ = 2--1 and $J$ = 5--4, and HCN, $J$= 1--0 and $J$ = 3--2 from QL07, convolving the brightness distribution with the beam of the single-dish telescope. 
In the next section we discuss the emitting regions assumed for \nueve\ and HCN within this new structure of the CSE around AFGL\,2343.
In the fitting process we only allow small changes in the properties found for the dense shell by CC07, while the parameters of the new component are free.

We have not tried to model our SO maps, due to their S/N ratio.
\subsection{Fitting results}

Our model assumes spherical symmetry and isotropical expansion. However, as said before, the observational maps show a maximum towards the east in the central channels and an intense peak at positive velocity. Since our model provides symmetrical maps with respect to the beam mayor axis (due to the convolution with the ellipsoidal beam), we adjust the parameters of our model to reproduce the average value of the two halves of the emission map.
Asymmetries in the profiles further than self-absorption cannot be modeled. 

{ We have assumed that the envelope structure is essentially that deduced by CC07 to explain 
the CO maps. As a result, we will see that our envelope model include regions that in fact are not probed by our SiO and HCN data. The properties of these regions are therefore given by our previous fitting of the CO maps.}

We have found that the parameters deduced for the densest part of the CO envelope by CC07 do not fit properly the molecular emission from HCN and \nueve. In particular, the strong emission of the high-$J$ transitions of HCN and $^{29}$SiO, compared to that of the lower-$J$ transitions, requires higher values of the density and temperature than those found for the CO lines. In the case of \nueve, once we fitted the profile and map of the $J$=2--1 transition with the conditions of density and temperature found for CO, the intensity predicted by the model for the $J$=5--4 transition is $\sim$20 times lower than the observed profile.
This supports the presence of the HE shell, as proposed in Sect.\,3.1, that must be hotter and denser than the CO dense shell found by CC07.
The \nueve\ emission maps show some characteristics that are different from the HCN maps, like the displacement of the maxima at the central velocities towards the center (Sect.\,3.2, Fig.\,5), and the fact that the asphericities in the central channels are more evident for \nueve\ than for HCN. This suggests that the emission from \nueve\ comes only from the HE shell while HCN also comes from the dense CO shell found by Castro-Carrizo et al. (2007). Anyhow, note that with the information that we have, we can not discard the presence of \nueve\ emission also coming from this dense shell.

The inclusion of this new HE component within the emission shell modifies in principle the predictions of the CO emission with respect to CC07. 
The CO emission is optically thick. { Therefore, a higher value of the density does not affect the previous CO fitting. On the contrary, a higher value of the temperature significantly raises the CO intensity. Due to this, the inclusion of the hot HE shell would result in an increase of the intensity with respect to previous calculations.}
This effect is more evident for the CO $J$= 2--1 transition than for the $J$= 1--0, so the predictions  obtained using parameters found by CC07 do not fit the CO maps once included the new HE component. To minimize this effect the HE region within the dense CO shell must be thin, such that this new emission would be mostly diluted by the convolution with the beam.  
This assumption about the width of the new component is confirmed by the excitation conditions required by the HCN data (see below).

{

   \begin{figure*}[t!hpb]
   \centering
   \includegraphics[width=6cm,angle=-90]{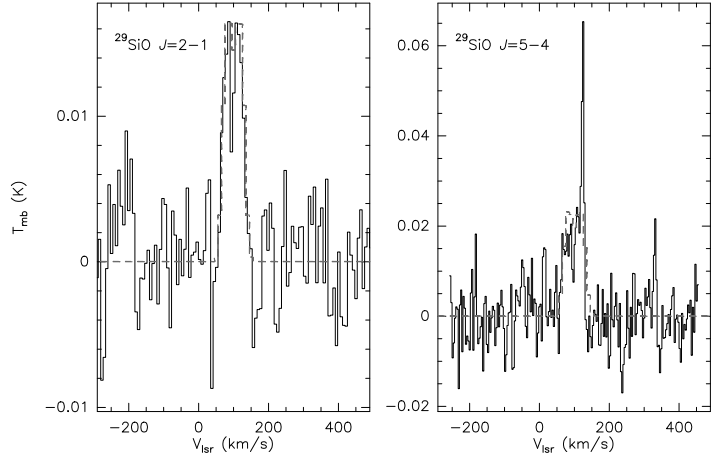}
      \caption{{
	Observed profiles from QL07 of \nueve\ $J$=2--1 ({\it left}) and $J$=5--4 ({\it right}) in solid lines 
compared with the synthetic profiles from the model in dashed lines.
              }}
         \label{model1}
   \end{figure*}
}
   \begin{figure*}[t!hpb]
   \centering
   \includegraphics[width=6cm,angle=-90]{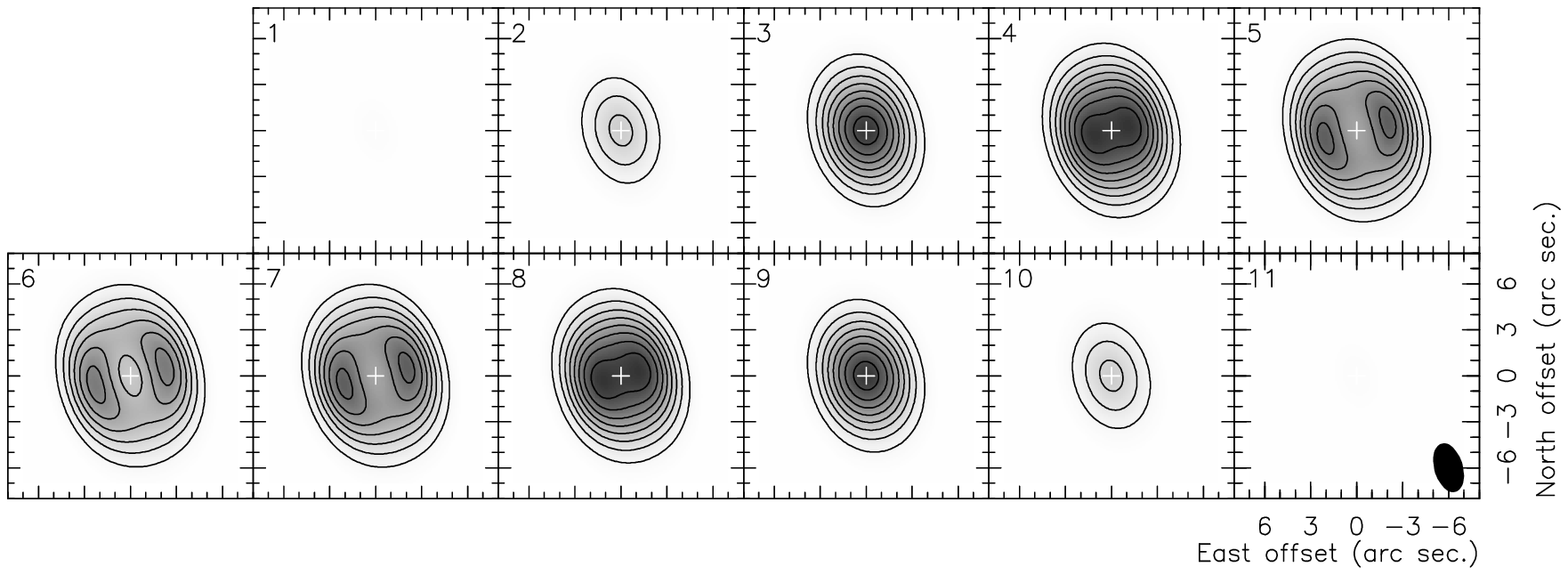}
      \caption{
Synthetic map obtained for the \nueve\ $J$= 2--1 emission in AFGL\,2343. The panels correspond to those of the observed maps (Fig.\,1) with the same level step. 
              }
         \label{model1}
   \end{figure*}
   \begin{figure*}
   \centering
   \includegraphics[width=8cm,angle=-90]{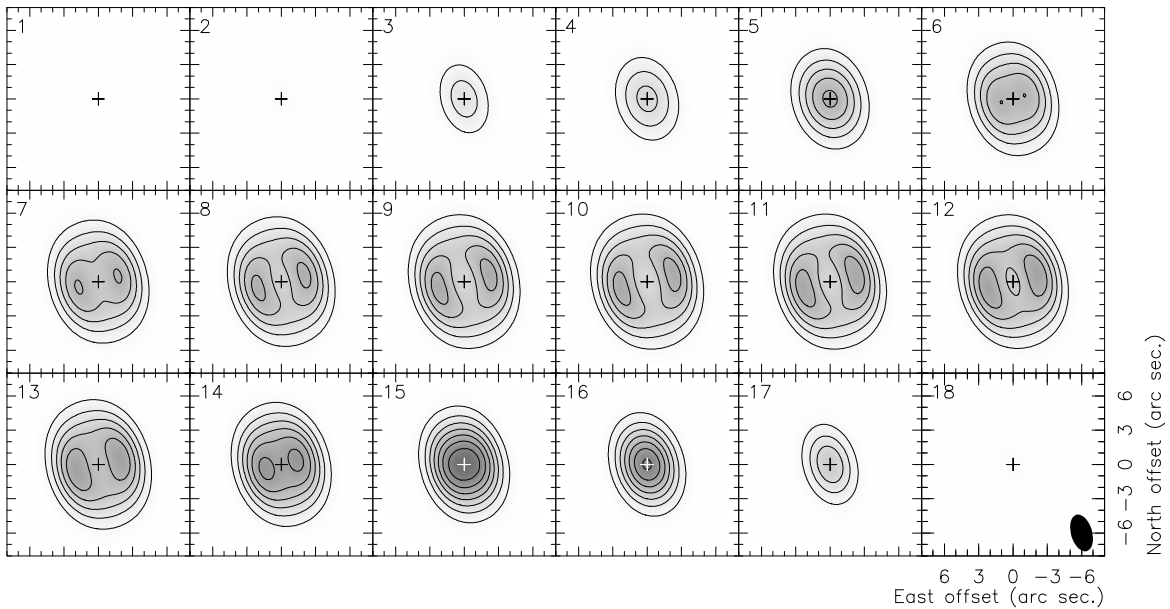}
      \caption{
Synthetic map obtained for the HCN $J$= 1--0 emission in AFGL\,2343. The panels correspond to those of the observed maps (Fig.\,2) with the same level step. 
              }
         \label{model2}
   \end{figure*}
\begin{figure*}
\centering
\includegraphics[width=5cm,angle=-90]{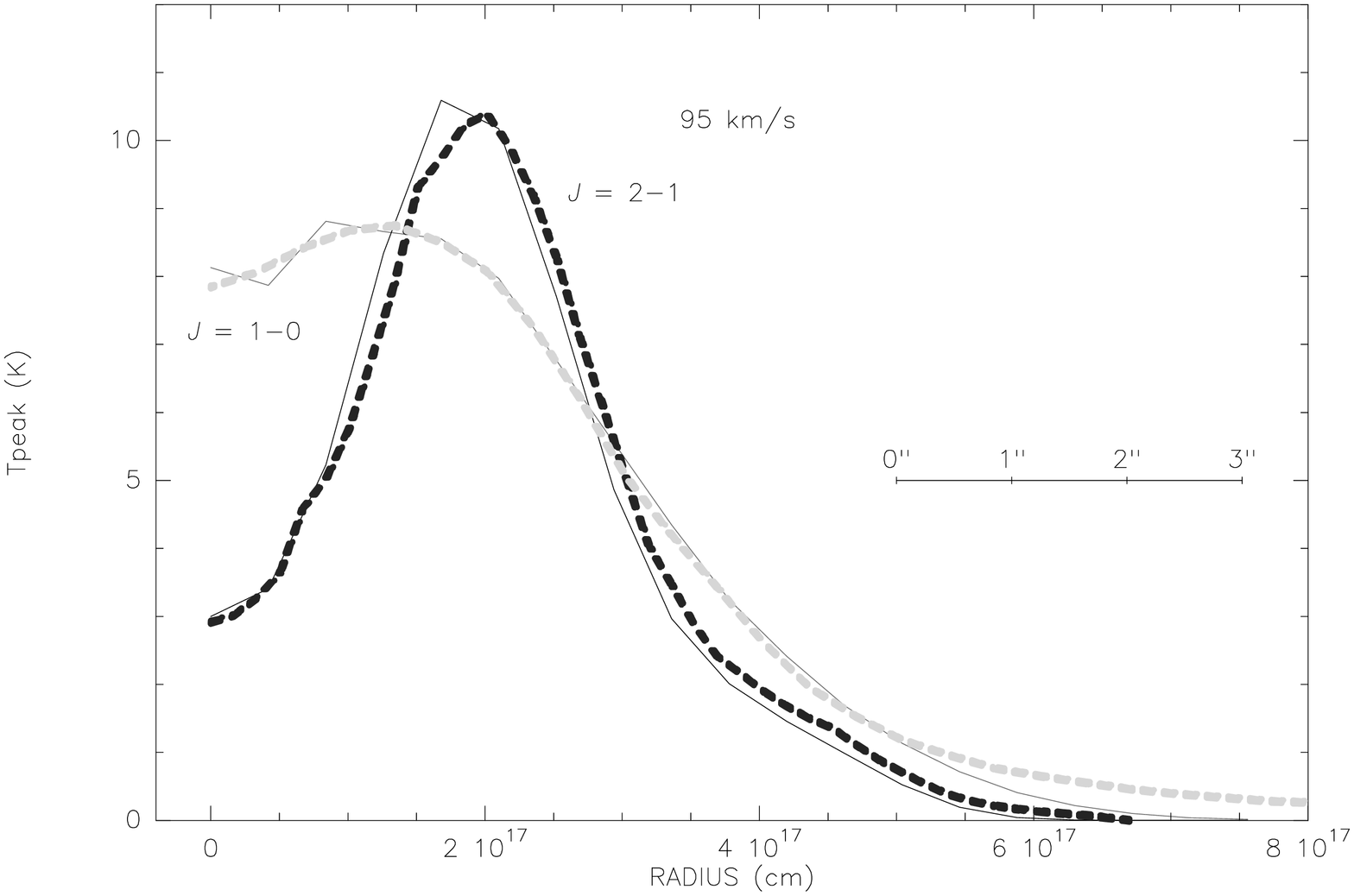}
\includegraphics[width=5cm,angle=-90]{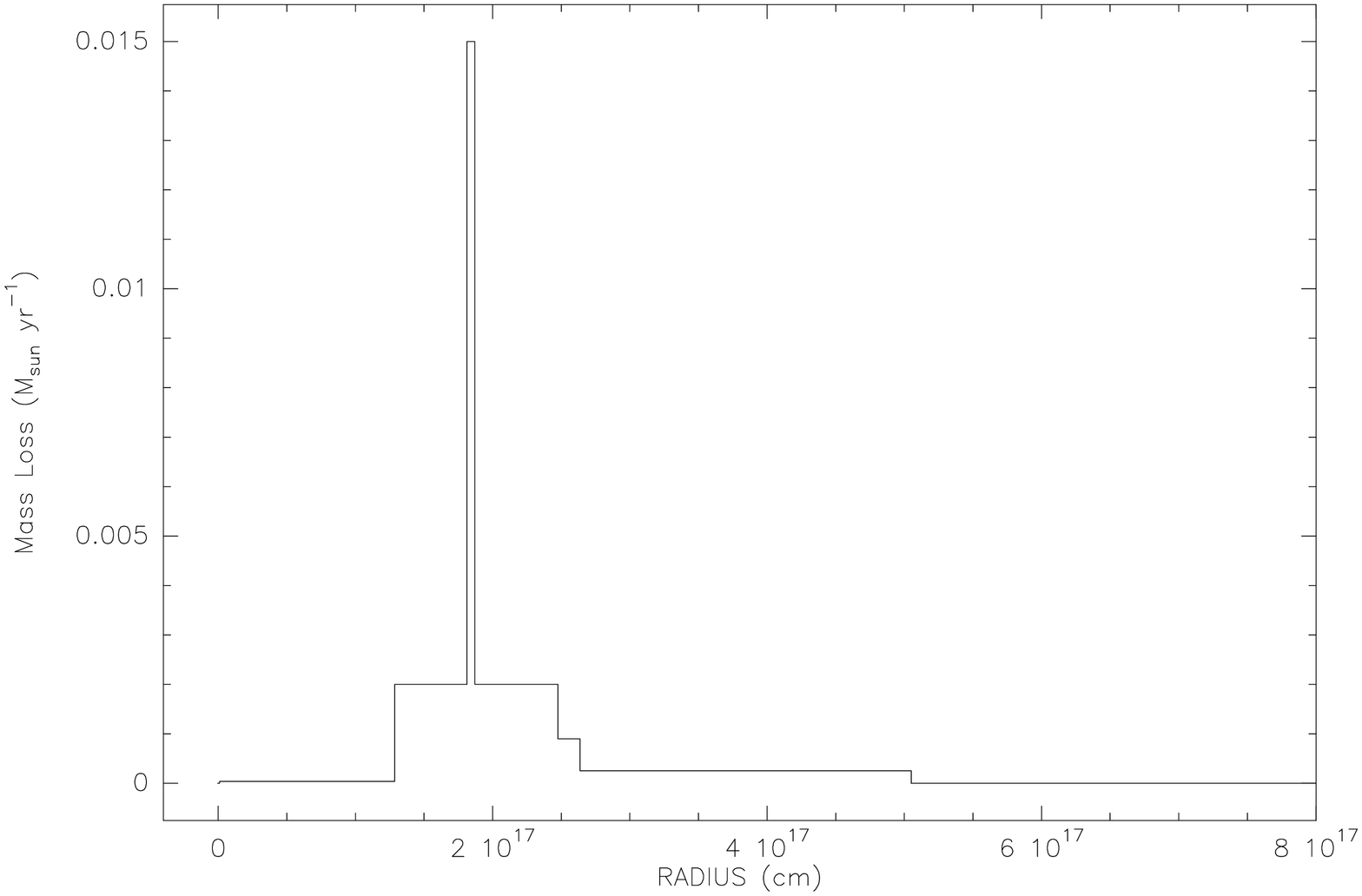}
\caption{$\it Left$: Azimuth-averaged brightness distribution (in dashed lines) compared with model results (solid lines) for AFGL\,2343, at $V_{\rm LSR}$ =
95 \kms. $^{12}$CO $J$ = 1--0 data are plotted in grey, $J$ = 2--1 data in black. $Right$: Mass-loss pattern found for AFGL\,2343.}
\label{masaCO}
\end{figure*}

\begin{table*}[t!hpb]
\caption{New properties of the CSE around AFGL\,2343 derived by fitting the maps of CO, \nueve\ and HCN and the profiles shown by QL07. { The value $T_{\rm min}$=8\,K}. The parameters not presented here but in CC07 show no significantly change. $^*$: New mass loss values.} 
\label{table:2}
\centering                          
\begin{tabular}{c c c c c c c c c c}
\hline\hline
Shell&R$_{\rm in}$ (cm)&R$_{\rm out}$ (cm)&\mloss(\ms yr$^{-1}$)& $T_{\rm 17}$(K)&$\alpha_{\rm t}$&X$_{\rm CO}$&X$_{\rm HCN}$&X$_{\rm ^{29}SiO}$&Molecules\\
\hline
1&1.1$\times$10$^{15}$&1.4$\times$10$^{17}$&4.3$\times$10$^{-5}$&40&0.5&3.0 10$^{-4}$&0&0&CO\\
2$^*$&1.4$\times$10$^{17}$&2.7$\times$10$^{17}$&2.1$\times$10$^{-3}$&13.4&0.5&3.0 10$^{-4}$&3.2 10$^{-8}$&0&CO,HCN\\
3&2.7$\times$10$^{17}$&2.8$\times$10$^{17}$&9.6$\times$10$^{-4}$&22.5&0.7&3.0 10$^{-4}$&0&0&CO\\
4$^*$&2.8$\times$10$^{17}$&5.4$\times$10$^{17}$&2.7$\times$10$^{-4}$&43&0.7&3.0 10$^{-4}$&0&0&CO\\
\hline
HE shell&1.93$\times$10$^{17}$&2.0$\times$10$^{17}$&1.6$\times$10$^{-2}$&31&0.5&3.0 10$^{-4}$&3.2 10$^{-8}$&2.2 10$^{-8}$&CO,HCN,\nueve\\
\hline

\end{tabular}
\end{table*}

We have accordingly modeled the emission of the CO and HCN maps and profiles shown in QL07 assuming that their emissions come from the dense shell found in CO by CC07 and a new HE component within it, hotter and denser. For \nueve\ we only take into account the HE component.

The resolution of the interferometric maps is not enough to determine accurately the size and the radii of the HE shell. Instead, we can determine a radius interval within which it lays by comparing our predictions with the observations. As we assume that \nueve\ emission only comes from the new component, this molecule is the tracer of the location and size of the new shell. 
We find that the \nueve\ observed maps are compatible with the emission of a shell located almost at the center of the CO dense shell, but slightly displaced towards the star, in order to explain the maxima found in the \nueve\ maps located in slightly inner regions than for CO maps (Fig.\,5).
The HE component in our model is located between { 1.93 10$^{17}\,$cm and 2 10$^{17}$\,cm}. 
A displacement of  {2.7 10$^{16}$\,cm} in the position of the layer starts to show a noticeable difference between the synthetic and the observed \nueve\ maps.
The values given for the width of this new component are indeed an upper limit
to the width of the HE region. 
{ The effects of the HE region on the emission of HCN mainly depend of the size of this region.} 
If the new component is too wide, { the emission from the high-excitation region will be dominant and, using the parameters of this component needed to fit \nueve\ $J$=5--4, we would obtain too intense  HCN $J$=3--2 profiles.} 
Also, { as said above,} a too wide shell would also significantly affect the CO emission 
{ due to its higher temperature.}
We have found that the width of this region must be smaller than  {2.1 10$^{16}$cm, in order to reproduce, within the uncertainties, the emission of all the different molecular transitions}. The position and width of the HE shell are therefore relatively well constrained by the data. 

The conditions derived from the fitting of all the data set (CO, \nueve, HCN maps and profiles presented in QL07) are presented in Table 1, { where $R_{\rm in}$ and $R_{\rm out}$ are the inner and outer radii of each shell, $T_{17}$ the temperature at $r=10^{17}$\,cm, and $X_{\rm Z}$ the relative abundance for molecule $Z$. 
The dependence of these parameters with the distance is: $M\propto D^2$,$\mloss\propto D$, $T(r)\propto D^{\alpha_{\rm t}}$, $R_i\propto D$.
{ The good fitting of the \nueve\ $J$=2--1 and $J$=5--4 profiles (Fig.\,6)}, only taking into account the HE shell, demonstrates the existence of a thin high-excitation shell, different from the wide envelope observed in CO.
The  synthetic maps of \nueve\ and HCN are presented in Figs.\,7 \& 8 respectively (in the electronic version). We recall that our model assumes spherical symmetry and isotropical expansion. Therefore, it yields symmetric profiles and maps, and can only fit averaged features, not reproducing the asymmetries of the SiO data. }

%

From this new modeling,
we obtain a total mass for the whole CO-rich envelope of { $M$=4.3\,\ms}. { The mass of the shell responsible for the emission of HCN is 3.4\,\ms, including the \nueve\ emitting shell (the new HE shell), which has a mass of 0.8\,\ms}. These values are lower than those previously derived by QL07 (a 9\% lower in the case of HCN and 4 times for \nueve) and, accordingly, the abundance values for these molecules increase by the same factor with respect to QL07 estimates. { Note that the masses presented by QL07 must be corrected by the distance to be compared with those given here.} The new mass-loss history and the new predictions for the CO emission are summarized in Fig.\,9.

{ As said in Sect.\,3.1, the new HE shell could correspond to a short phase of very high mass-loss rate or to shock effects, in view of its relatively high density, temperature, and the SiO abundance. In the second case, Fig.\,9 (right panel) would not indicate that this shell is due to an increase of the mass-loss rate, but just show a relative increase of the density due to the interaction of two shells.}

The \nueve\ abundance obtained here is compatible with that obtained by QL07, once the different mass values are taken into account.
On the contrary, the abundance deduced from our detailed study for HCN is higher than that found by QL07, even if this correction is introduced. 
This is partially due to the simplications in the treatment of the rotational levels and opacities made by QL07. 
Although the method by QL07 was suitable for moderate values of the opacity, the values we found for the optical depth of the HCN lines are { too} high to be well treated in that way.
The values of the relative abundances found here are { 3.2 10$^{-8}$ for HCN and 2.2 10$^{-8}$} for $^{29}$SiO (HE shell), see Tables 1 \& 2. 

\section{Molecular abundances} 

The abundance calculation by QL07 was carried out with some simplifications in the treatment of the excitation of the rotational levels, as well as assumptions on the extent of the emitting region for the molecular emission, which we have found to be inaccurate. In particular, we have found here that the values of the mass emitting in HCN and \nueve\ are smaller than those assumed in QL07.

We find higher values for the molecular abundances in the respective emitting regions than those found by QL07, as we can see in Table 2 and previous section. Anyhow, this increase in the abundances does not solve the underabundance problem mentioned by QL07, it only makes it less severe. 

\begin{table}
\renewcommand{\footnoterule}{}
\label{table:1}
\centering
\caption{New mean molecular abundances derived for 
AFGL\,2343. For $^{12}$CO we adopt the same abundance that assumed by CC07. For details of the emitting regions, see Table.\,1.}
\scalebox{1}{

\begin{tabular} {l c c}
\hline\hline
Molecule&	$\langle\,X\,\rangle$&Emitting regions\\
\hline
$^{12}$CO	&3.0\,10$^{-4}$&1 + 2 + 3 + 4 + HE shell.\\
$^{13}$CO	&4.0\,10$^{-5}$&1 + 2 + 3 + 4 + HE shell.\\
C$^{18}$O	&{3.8\,10$^{-8}$}&1 + 2 + 3 + 4 + HE shell.\\
HCN		&{3.2\,10$^{-8}$}&2 + HE shell.\\
HNC		&{3.5\,10$^{-9}$}&2 + HE shell.\\
SiO		&{4.0\,10$^{-7}$}&HE shell.\\
$^{29}$SiO	&{2.2\,10$^{-8}$}&HE shell.\\
\hline
\end{tabular}

}

\end{table}

With the new characteristics found for the CSE around AFGL\,2343 we are able to compute the abundances of some of the other molecules observed by QL07, using the line profiles published by these authors. We have calculated the new abundance values for $^{13}$CO, C$^{18}$O, HNC and SiO (see Table 2). The profiles from other molecules, like CS, show peculiarities that can not be easily interpreted with the model derived here. For example, the line profiles are particularly thin and there is no profile emission peak at positive velocities. Therefore, the abundances of these species are not discussed again.

For the new estimates, we have taken the collisional coefficients from Flower (2001) for $^{13}$CO and C$^{18}$O, Turner et al. (1992) for SiO, and adopted the same coefficients for HNC as for HCN; all of them obtained from the LAMDA database (see Sect.\,4.1 for details). We have taken different emission regions for the different molecules, according to the observed spectral features, by comparison with the properties of the emissions of \doce, \nueve, and HCN, which have been well studied. The emission from CO and its isotopes would come from all the envelope described in Table 1, the emission of SiO and isotopes is assumed to come from the new HE shell, and for HCN and HNC we assume that their emission comes from the densest shells (shell 2 and the HE shell in Table 1) of the envelope. The results for the new computed abundances are shown in Table 2.

The deduced abundances are in most of the cases higher than those found by QL07. This is due, as said above, to a better description of the rotational levels, a better treatment of the opacity, and the differences in the emitting masses. The abundance obtained for C$^{18}$O, on the contrary, is low compared with previous calculations, because the excitation temperature estimated in by QL07 was high compared with those obtained using the LVG code. However, the abundances still remain low with respect to the values found for the YHG IRC+10420 and the O-rich AGB stars.


{ Most of the molecules detected in AFGL\,2343 by QL07 are expected to be dissociated at these large 
distances (1.4 10$^{17}$ -- 2.7 10$^{17}$\,cm). The massive circumstellar medium may partially shield the UV radiation, explaining the abundances found for AFGL\,2343. In the case of IRC\,+10420, although the mass derived for the molecular gas is lower than for AFGL\,2343, the region from which the molecular emission comes is closer to the star. At these radii the abundances are probably unaffected by photodissociation, leading to higher abundances in IRC\,+10420 than in AFGL\,2343. Note that the molecular abundances in IRC\,+10420 are often similar to those of O-rich AGB stars. In the case of SiO and its isotopes, their abundance in AGB circumstellar envelopes is known to be high only in the very inner shells. In fact, its presence further from central compact regions is thought to be associated with shocks (Sect.\,3.1): a shock front may heat the grains where SiO is depleted evaporating this molecule from them and increasing its abundance in the gas. }

\section{Conclusions}

We have discovered the presence of a high-excitation region (HE region) within the densest shell found from CO maps by Castro-Carrizo et al. (2007; CC07). Although this region is not directly detected (due to the limited resolution), its characteristics can be inferred from the maps and line profiles of HCN and $^{29}$SiO, as shown in previous sections. The presence of this high-excitation shell is in particular indicated by the emission of high-$J$ lines compared with lower rotational transitions (see line profiles in Quintana-Lacaci 2007; QL07). Also, an emission peak is clearly seen in the interferometric maps of HCN and \nueve, very remarkable for high-$J$ transitions. This maximum is also seen in the obtained maps at positive velocity (see Sect. 3.1). 
These results can be understood as showing the existence of an inhomogeneous shell expanding at higher velocity and with higher excitation than the rest of the envelope, 
leading to the asymmetries found in velocity and in brightness distribution. All these features were not very noticeable in the CO lines, probably because they do not require high excitation. 
The characteristics derived for this shell (namely: it is relatively thin, dense, hot, and rich in SiO) suggest that this component can { be associated  to a shock process}. { However, it is also possible that this HE shell is the result of a period of particularly high mass-loss rate. The time required to form the HE shell found here, see Table 1, is just 50 years. Another YHG, $\rho$ Cas, is known to be passing through similar periods of high mass loss. Lobel et al. (2003) derived a mass-loss rate of $\sim$5.4 10$^{-2}$\my for the outburst of $\rho$ Cas in 2000-2001. In particular, similar intense outbursts have been detected on $\rho$ Cas 
three times in the last century, each of them lasting several hundreds days.
Therefore, such brief, high mass-loss periods are not unlike to occur in YHGs. If the HE shell of AFGL\,2343 is the result of these outbursts, these events occurred $\sim$2000 years ago. Since CO rotational emission has not been detected in $\rho$ Cas, we can interpret that $\rho$ Cas did not underwent similar process in the past and, therefore, it is in an earlier evolutionary stage than AFGL\,2343.}

A thin shell rich in SiO was also found in \irc\ by Castro-Carrizo et al. (2001). In \irc\ the SiO emission was found to come only from a thin shell located in the inner part of the dense CO shell. These authors also suggested that this SiO shell could be related with a shock. 
Therefore, the SiO shell found in \irc\ and the high-density shell found here for \afg\ can be a common consequence of the episodic mass loss processes that occurs in this type of stars along their evolution. These sources are thought to have periods of enhanced mass loss, as shown by theoretical considerations (e.g. Nieuwenhuijzen \& de Jager, 1995) and by observations, like the recently quasi-explosive mass ejection period ($\mloss\sim6 10^{-2}$) observed for $\rho$ Cas (Lobel et al. 2003) or by the mass loss pattern found by Castro-Carrizo et al. (2007) for IRC\,+10420 and AFGL\,2343. See also the mass loss pattern that we have obtained in Fig.\,9.

The presence of a new shell modifies the molecular abundance estimate with respect to QL07.
We have fitted our maps of \nueve\ and HCN, as well as the profiles of some molecules presented by those authors, using the model described in Sect. 5, and obtained new values of the abundances (see Table 2). In the new abundance estimates, we have taken into account the envelope structure presented in this paper and we have more accurately taken into account the opacity effects and level excitation.

\acknowledgements
{This work has been supported by the Spanish Ministerio de Ciencia y
Tecnologia and European FEDER funds, under grants AYA2003-7584 and 
ESP2003-04957.}

{}

\end{document}